\documentclass[manuscript]{aastex}

\usepackage{graphicx}
\usepackage{natbib}
\usepackage{color}
\fontsize{6.5pt}{0pt}\selectfont

\slugcomment{Not to appear in Nonlearned J., 45.}

\shortauthors{Oba et al.}

\begin{document}
\title{The small-scale structure of photospheric convection retrieved by a deconvolution technique applied to \textit{Hinode}/SP data}

\author{T. Oba\altaffilmark{1,2}}

\author{T.L. Riethm{\"u}ller \altaffilmark{3}}

\author{S. K. Solanki \altaffilmark{3,4}}

\author{Y. Iida\altaffilmark{5}}

\author{C. Quintero Noda,\altaffilmark{2}}

\author{T. Shimizu\altaffilmark{2}}

\altaffiltext{1}{SOKENDAI (The Graduate University for Advanced Studies), 3-1-1 Yoshinodai, Chuo-ku, Sagamihara, Kanagawa 252-5210, Japan }
\altaffiltext{2}{Institute of Space and Astronautical Science, Japan Aerospace Exploration
Agency, 3-1-1 Yoshinodai, Chuo-ku, Sagamihara, Kanagawa 252-5210, Japan }
\altaffiltext{3}{Max-Planck-Institut f{\"u}r Sonnensystemforschung (MPS), Justus-von-Liebig-Weg 3, 37077 G{\"o}ttingen, Germany}
\altaffiltext{4}{School of Space Research, Kyung Hee University, Yongin, 446-701 Gyeonggi, Republic of Korea}
\altaffiltext{5}{Department of Science and Technology/Kwansei Gakuin University, Gakuen 2-1, Sanda, Hyogo, 669-1337 Japan}

\begin{abstract}
\footnotesize{Solar granules are bright patterns surrounded by dark channels called intergranular lanes in the solar
photosphere and are a manifestation of overshooting convection. Observational studies generally find stronger
upflows in granules and weaker downflows in intergranular lanes. This trend is, however, inconsistent
with the results of numerical simulations in which downflows are stronger than upflows through
the joint action of gravitational acceleration/deceleration and pressure gradients. One cause of this
discrepancy is the image degradation caused by optical distortion and light diffraction and scattering that takes place in an
imaging instrument.
We apply a deconvolution technique to \textit{Hinode}/SP data in an attempt to recover the original solar scene.
Our results show a significant enhancement in both, the
convective upflows and downflows, but particularly for the latter. 
After deconvolution, the up- and downflows reach maximum amplitudes of -3.0 km/s and +3.0 km/s at an average geometrical height of roughly 50 km, respectively. 
We found that the velocity distributions after deconvolution 
match those derived from numerical simulations. 
After deconvolution the net LOS velocity averaged over the whole FOV lies close to zero as expected in a rough sense from mass balance.  }
\end{abstract}

\section{Introduction}
\footnotesize{Solar granules are bright patches in the solar photosphere surrounded by dark channels called intergranular lanes. 
The granules are a manifestation of overshooting hot upflows from the convectively unstable subsurface layers into the stable photosphere \citep{Stix2004}. 
There, the excess pressure above the upflows accelerates the gas sideways. 
Radiative cooling, along with an associated increase in density, makes the material both darker and less buoyant, and the material finally starts to flow down, where it collides with horizontally flowing gas from a neighbouring granule. A dark intergranular feature is formed at the location of the downflow. \\
　In the past 2 decades, three-dimensional numerical simulations have progressed enough to reproduce many properties of the solar atmosphere, providing physical insights into, and leading to a better understanding of, solar convection (e.g., \citealt{Nordlund2009}). 
Simulated granulation (\citealt{Stein1998}) shows a typical pattern with slower upflowing gas, and with faster downflows, e.g., \citet{Hurlburt1984}. 
However, so far observational results do not support downward motions faster than upflows. 
\citet{Hirzberger2002} investigated the velocity field of the photosphere and reported almost the same magnitude of 1.5 km/s for both, upflows and downflows. 
\citet{Kostik2007} used a bisector analysis to extract the height structure of the velocity field from the asymmetric shape of 
an absorption line and also found that both upward and downward motions have approximately the same amplitude at heights from 0 km to 500 km. 
By avoiding image degradation due to atmospheric seeing and with its precisely known PSF, the \textit{Hinode} spacecraft \citep{Kosugi2007} is well suited to precisely derive the velocity field. 
Using the spectrum observed by the Solar Optical Telescope (SOT) \citep{Tsuneta2008} onboard \textit{Hinode}, \citet{Yu2011b} found that the velocity field ranged from $-$2.0 km/s (upward) to less than +2.0 km/s (downward) and the peak of their distribution coincided with upflows. 
Using a Milne-Eddington inversion code, \citet{Jin2009} found a velocity field dominated even more strongly by blueshifts with a range from $-$3.3 km/s to +2.0 km/s and a peak at -1.0 km/s. 
Finally, using a bisector analysis, \citet{Oba2017} investigated the average convective velocity field of upflows and downflows separately. 
They found 1.5 times stronger upward than downward motions at photospheric heights from 40 to 160 km. \\
　One possible cause of this discrepancy between the numerical and observational results is the imaging performance of the employed telescopes. 
\citet{Danilovic2008} reported the significance of this effect based on a careful study in which they applied the point-spread function (PSF) of the instrumental configuration of the \textit{Hinode}/SP (the spectropolarimeter on \textit{Hinode}) on 3D magneto-hydrodynamic (MHD) simulations (cf. \citealt{Wedemeyer2008}, who derived the \textit{Hinode}/SOT PSF empirically). 
They blurred the continuum intensity map obtained by running the SPINOR spectral synthesis code \citep{Frutiger2000} on a MHD simulation snapshot produced by the MURaM code \citep{Vogler2005}. 
The degradation of the data reduced the RMS intensity contrast from 14.4\% to 7.5\%.
The latter value is nearly the same as its observational counterpart of 7.0\%.
This difference in contrast of a factor of 2 suggests that the ratio between the velocities in the upflows and downflows may also be significantly affected by the PSF of the \textit{Hinode}/SP. 
Therefore, a removal of the degradation owing to the PSF could help restore the velocities to values closer to the solar ones. 
Testing this hypothesis is the main aim of this study. \\
　We attempted to properly capture a better approximation of the convective velocity signal through a spatial deconvolution process.
Recently, some deconvolution analyses have been performed on the \textit{Hinode}/SP data. 
This instrument is well-suited for such an image reconstruction technique because its PSF is well-known and almost time-independent (except for focus changes). 
Early work was done by \citet{Mathew2009} and \citet{Wedemeyer2009}, who investigated the intensity contrast in the photosphere. 
A self-consistent approach was developed by \citet{vanNoort2012} and \citet{vanNoort2013}, who extended the SPINOR code to iteratively invert a set of profiles while also taking account of the influence of the PSF. 
The resulting technique was applied to investigate active regions (cf. \citealt{Riethmuller2013}; \citealt{Tiwari2013}; \citealt{Buehler2015}, etc.) and the quiet Sun \citep{Danilovic2016}. 
To precisely derive the photospheric physical parameters in active regions using the SIR inversion code (\citealt{RuizCobo1992}), \citet{Ruizcobo2013} and \citet{Quintero2015} deconvolved spectroscopic scans with the known PSF of \textit{Hinode}/SP.
They countered the enhancement of noise, typical for deconvolution techniques, by suppressing the noise through a principal component analysis (PCA).
To these established approaches (\citealt{vanNoort2012} and \citealt{Ruizcobo2013}), we add a simple alternate deconvolution method combining noise suppression with deconvolution and investigate how its application affects the up- and downflows due to the photospheric convection. \\
　We carry out our analysis in two main steps. First we evaluate our deconvolution technique. 
To determine how well the deconvolution recaptures the original image, we use synthesized maps created by numerical radiation MHD simulations that were degraded by the \textit{Hinode} PSF. 
In a second step, the vertical velocity in solar granulation is determined more reliably than in previous analyses. 
This observational result is obtained by applying the deconvolution technique to a time series of spatial scans carried out by the \textit{Hinode}/SP. \\
　This paper is structured as follows. 
In Section 2, we describe the observations, numerical simulations, and our approach to deriving the convective motion. 
In Section 3, we evaluate the deconvolution technique and observational results. 
In Section 4, we discuss how the photospheric convection changes through the deconvolution. 
Finally, in Section 5, we summarize our findings. \\

\section{Methodology}

\subsection{Observations}
The observations analyzed in this work were taken by the Spectropolarimeter ($SP$) \citep{Lites2013b} 
on the SOT on board the \textit{Hinode} spacecraft \citep{Kosugi2007,Tsuneta2008,Suematsu2008,Shimizu2008} on August 25, 2009, between 08:01 and 09:59 UT. The \textit{Hinode}/SP observed a quiet region at solar disk center, recording the Stokes $I$, $Q$, $U$, and $V$ profiles of the Fe~{\sc i} 630.15~nm and 630.25~nm spectral lines. It used a spectral sampling of 21.5~m\AA, and a pixel size of approximately 0.$^{\prime \prime}$16, providing a spatial resolution of 0.$^{\prime \prime}$32. The integration time was 1.6 s per slit position. The slit had a length of 60.$^{\prime \prime}$9 and was oriented in the solar N--S direction. 
A small field-of-view (FOV) of $4.^{\prime \prime}5\times60.^{\prime \prime}9$ was scanned 117 times with 30 slit positions per scan (each slit step corresponded to 0.$^{\prime \prime}$15) and a cadence of approximately 62~s. Fig.~\ref{fig:1} shows the continuum intensity for the first 11 scans where the granulation pattern and its evolution can be clearly identified. 
The SP data were calibrated with the standard routine SP\_PREP in the Solar SoftWare package (\citealt{Lites2013a}). 
Finally, the velocity was calibrated as in \cite{Oba2017}. \\
　\textit{Hinode}/SP obtains spectral images by slit scanning. Consequently, each slit position along the E--W direction is recorded at a different time. 
For the precise derivation of the convective velocity field, we wanted to create a two-dimensional spectral map where all spatial locations corresponded to the same solar time because the deconvolution process needs the information outside the observed slit position. 
Such a uni-temporal image is also needed to remove the 5-min oscillations, which are larger scale phenomena producing fluctuations in both horizontal directions at the solar surface. 
Accordingly, we performed a linear time interpolation using the information from consecutive raster scans to produce a pseudo-instantaneous spatial image (see more details in Fig.\ref{fig:20} and its caption). 
Since the time interval of linear interpolation (roughly 1 min in this case) is shorter than the typical 6-min granulation lifetime \citep{Hirzberger1999} and the 5 min period of p-modes, this interpolation should not unduly influence our results in unwanted ways. 

\subsection{Synthesis of Stokes profiles}
　Before using the deconvolution technique developed in this work on \textit{Hinode}/SP observations we wanted to first test it on synthetic observations to verify its possibilities and limitations. 
For this purpose, we needed to synthesize the Stokes profiles, solving the radiative transfer equation starting from a given atmosphere. This was performed using the numerical code SPINOR (\citealt{Frutiger2000,Frutiger2000b}) that works under the assumption of local thermodynamic equilibrium, which is adequate for the Fe~{\sc i} lines at 630~nm. \\ 
　The input atmospheres used by SPINOR were a series of snapshots taken from a rediative MHD simulation generated by the  MURaM code \citep{Vogler2005}. 
The chosen setup was the same as that used in \cite{Riethmuller2014}. 
The calculation ran for 1 hour of solar time and we stored a snapshot containing all the physical parameters (needed for the line computation) every 30 seconds, thus producing 120 snapshots in all. 
Each snapshot had a size of 6 $\times$ 6~Mm with a spatial sampling in the horizontal directions of 10.42~km. 
The vertical extent of the simulated domain was 1.4~Mm and the grid size was constant at 14~km. The average height where the continuum optical depth ($\tau_{500}$) was unity was reached 0.5~Mm below the top boundary of the computational box. 
The simulation was performed without any initial magnetic flux, and therefore, we focused only on the pure hydrodynamics in the photosphere. \\
　Figure \ref{fig:21} shows an example of the spatial distribution of the intensity signals at the continuum wavelength $\lambda=630.1$~nm for a selected snapshot.
 In addition, we can see that the contrast is large (with the RMS being 16.8\%), as we had not yet applied any type of spatial degradation to these images.

\subsection{Spatial deconvolution}\label{bozomath}

In this section, we explain the basis of the deconvolution technique. The observed image can be expressed as
\begin{equation}
I\left( x,y\right) =\int \int ^{\infty }_{-\infty }I_{0}\left( \xi ,\eta \right) PSF\left( x,y:\xi ,\eta \right) d\xi d\eta + \epsilon \ (x, y), 
\label{eq:deconv}
\end{equation}
where $I\left( x,y\right)$ is the observed intensity, $I_{0}\left( \xi ,\eta \right)$ is the ideal image, $PSF\left( x,y:\xi ,\eta \right)$ is the point-spread function, and $\epsilon$ is random noise (the equation is often abbreviated in the form $I=I_{0} \ast PSF+\epsilon$, where asterisk means convolution ).
This equation signifies that the observed intensity at a given pixel results from the contribution by the surrounding pixels in the ideal image. 
The goal of deconvolution is to obtain the intensity at that pixel in the ideal image $I_{0}$, using a given $PSF$ and the observed intensity $I\left( x,y\right)$. \\

\subsubsection{Richardson-Lucy algorithm with regularization}\label{bozomath}
　Several types of spatial deconvolution techniques have been proposed (see for instance \citealt{Jansson1997} for an overview). 
In this study, we applied the Richardson-Lucy (RL) algorithm as a deconvolution method (\citealt{Richardson1972}; \citealt{Lucy1974}), which was also used in the previous studies involving deconvolution of \textit{Hinode} data (\citealt{Ruizcobo2013}; \citealt{Quintero2015}). 
This algorithm is an iterative scheme used to increase the probability of being Poisson distributed.  
We assumed the following image formation model: 
\begin{equation}
i(X)=o \ast h, 
\label{eq:eq1}
\end{equation}
where $i$ is the image, the variable $X$ is two dimensional and represents $(x, y)$, $o$ is the object, and $h$ is the $PSF$. 
Under the influence of Poisson noise, we have to deal with $i=\varphi (o \ast h)$, $\varphi$ being a Poisson distribution. 
The probability of our measurements falling within a Poisson distribution can be expressed as
\begin{equation}
p\left(  i | o\right) =\prod _{X}\left( \frac {\left[ h \ast o\left( X\right) \right] ^{i\left( X\right) }e^{-\left( h \ast o \right) \left( X\right) }}{i\left( X\right) !}\right),
\label{eq:dop}
\end{equation}
where ! is the factorial. 
Our goal was to find the $o(X)$ that maximized the probability of $p\left(  i | o\right)$ to most likely produce the observed value of $i(X)$. To search for the solution, we used the following iterative algorithm, 
\begin{equation}
o_{k+1}\left( X\right) =\left\{ \left[ \frac {i\left( X\right) }{\left( o_{k}\ast h\right) \left( X\right) }\right] \ast h\left( -X\right) \right\} o_{k}\left( X\right).
\label{eq:rl0}
\end{equation}
Equation (4) describes an iteration step of the RL algorithm that retrieves the real image through an iterative scheme. However, in the case of an observation with noise, the RL algorithm does not always converge to a suitable solution because of the amplification of the noise after several iterations. 
\citet{Ruizcobo2013} and \citet{Quintero2015} reduced the noise on the profiles  by the PCA (\citealt{Loeve1955}) before deconvolution. This involved creating a base of eigenvectors that included the spectral information of the original data set. \\
　We followed a different approach on treating the noise: we added a regularization term, so-called Tikhonov-Miller regularization, to reach a suitable solution \citep{dey2004}. 
This method searches for a spatially-smoothed solution in the ideal image by allowing non-maximum probability in finding the answer, by limiting the enhancement of fluctuations at the highest spatial frequency. 
Instead of the original Richardson-Lucy algorithm given in Equation (\ref{eq:rl0}), we obtained a regularized form of the RL algorithm,
\begin{equation}
o_{k+1}\left( X\right) =\left\{ \left[ \frac {i\left( X\right) }{\left( o_{k}\ast h\right) \left( X\right) }\right] \ast h\left( -X\right) \right\} \frac {o_{k}\left( X\right) }{1+2\lambda _{TM}\Delta o_{k}\left( X\right) },
\label{eq:rl}
\end{equation}
where $\lambda _{TM}$ is the regularization parameter, and $\Delta$ is the Laplacian operator. The observed image $o$ and the PSF $h$ are known parameters, whereas the regularization parameter $\lambda_{TM}$ can be chosen freely. It is important to set an appropriate value of $\lambda_{TM}$ because the success of the regularization depends on the amount of the noise to be suppressed; the larger the value of the parameter $\lambda_{TM}$, the greater the likelihood of a spatially-smoothed image with a large departure from the observed image; however, for a smaller value, the algorithm regards any observed signal (including the noise component) as a true one, so that it increases noise considerably, which is not desirable for a noisy image. The selection of the best parameter value is explained in Section 3.1.\\
　In this study we refrained from correcting for spectral degradation caused by the dispersion in the grating because the deformation of spectral profiles was considered to be negligible. Using a tunable laser, \cite{Lites2013b} measured the SP-spectral-response profile prior to launch and reported the difference of intensity before and after the convolution. They found that the line core showed the largest deviation of about 3\% in the absorption line. Our bisector analysis (see Section 2.4) did not make use of the line core, but only the intensity range from 0.45 to 0.70, so that the spectral degradation likely had little effect on the deduced velocity fields.

\subsubsection{PSF}\label{bozomath}
Our spatial deconvolution code, introduced in the previous section, requires a spatial PSF (see Eq.\ref{eq:rl}). 
We used the instrument-PSF data previously obtained from the \textit{Hinode} pupil presented in \citet{Suematsu2008} (for more details, see \citealt{Danilovic2008}). 
They modeled the optical path of the telescope and transfer optics to the entrance spectrograph slit, based on the commercial optical design software ZEMAX. 
Their computation included blurring effects due to the telescope aperture, central obscuration by the secondary mirror and its attachment spider, spatial sampling of the CCD detector, and defocus. 
The PSF function, sampled at 0.$^{\prime \prime}$16 ($\approx$100 km), which is the same pixel sampling as the dataset of the \textit{Hinode}/SP, is plotted in Fig.\ref{fig:2}. 
Since the scanned area is narrow, we need to limit the extent of the PSF. 
To this end, we calculated the total energy contained within a set of squares of different areas, each corresponding to a square number of detector pixels, 1$\times$1, 2$\times$2, 3$\times$3, etc.  The results are plotted in Fig.\ref{fig:3}. 
In raw data (i.e., without any deconvolution), any given pixel contains only 13\% of the observed intensity from the corresponding location, while the remaining 87\% originates from outside. 
The total energy contained in a given area increases rapidly as the area increases (Fig.\ref{fig:3}). 
We found that a square of 13$\times$13 pixels contained 96\% of the intensity. Assuming that the remaining 4\% does not significantly contribute to the observed information, we use only the part of the PSF contained within these 13$\times$13 pixels. To allow for a proper deconvolution, we normalized the PSF covering a square of 13$\times$13 pixels to have a total value of 100\%.

\subsection{Derivation of the convective velocity field}
　To derive the convective velocity field, we processed the deconvolved spectrum in two steps. First, bisector analysis provided the velocity field at multiple heights \citep{Stebbins1987, Dravins1981}, and then a subsonic filter \citep{Title1989} removed the 5-min oscillations to obtain the convective velocities.
Through these steps, we obtained convective velocity fields at six bisector levels (intensity level of $I/I_{c}$=0.70, 0.65, 0.60, 0.55, 0.50, 0.45) using the Fe~{\sc i} 630.15~nm spectral line (the details are described in \citealt{Oba2017}). 
We avoided sampling other intensity levels (more than 0.70 or less than 0.45), to make sure that our results are not adversely affected by pixels lost in intergranular lanes owing to their lower continuum and higher line core intensity, e.g., if a value of $I/I_{c}$ of 0.90 is chosen, then the bisector at this $I/I_{c}$ value cannot be determined in all the pixels at which the continuum intensity drops below  this value. This happens regularly in the intergranular lanes. The range of $I/I_{c}$ values are chosen such that the bisectors at these $I/I_{c}$ values can be determined in almost all spatial pixels. 
In a simple 1D model one can formally assign a geometrical height to an intensity level, i.e. to a bisector level.  
Based on the mean temperature stratification, those six bisector levels correspond to geometrical heights of 49, 62, 77, 92, 112, and 135 km, respectively. 
We stress, however, that as the solar atmosphere is highly structured, the actual height at which radiation is emitted changes strongly from one horizontal position to another. Such changes in height are often considerably larger than the whole height range formally covered by the bisectors (as given above). Therefore, these heights should be used with extreme caution.

\section{Results}
\subsection{Evaluation of the deconvolution technique using the synthesized image} 
To estimate how accurately our deconvolution technique removed the degradation by the instrument PSF and restored the observed images, we used images synthesized from a numerical simulation as test images. 
The advantage of this approach is that the ideal image (the simulated image) is known, and has a similar morphology and intensity contrast as the Sun \citep{Danilovic2008}. 
Recall that the forward calculation of the SPINOR code provided spectral data covering the wavelength range from 630.1 to 630.3 nm based on the atmospheric parameters simulated by the MURaM code (one snapshot of the continuum intensity is shown in Fig.\ref{fig:21}). 
This simulated continuum intensity map was binned to correspond to a spatial sampling of 100 km for a direct comparison with the \textit{Hinode}/SP observations. 
Hereafter we call it \textit{the answer-image}. 
\textit{The answer-image} obtained from the snapshot displayed in Fig.\ref{fig:21} is shown in Fig.\ref{fig:5} (a). 
\textit{The answer-image} is free of noise. 
To mimic the observed intensity map of the \textit{Hinode}/SP, we then convolved \textit{the answer-image} with the instrument PSF and introduced Gaussian random noise. 
We call the resulting, degraded image the \textit{imitated-image}. 
An example is shown in Fig.\ref{fig:5} (b). 
The variance of the introduced Gaussian random noise was calculated from the observed continuum intensity, and is 0.012. 
The retrieved image through the RL algorithm without any regularization after 50 iterations is shown in Fig.\ref{fig:5} (c).
The scatter plot of this image vs. the answer image is drawn in Fig.\ref{fig:5} (e) with a standard deviation of 0.0406. 
An image retrieved by the extended RL algorithm with regularization ($\lambda_{TM}=0.0016$ is adopted because this is the optimum value as explained later) is shown in Fig.\ref{fig:5} (d). 
A scatter plot of the image deconvolved including regularization of the deconvolved image versus the intensities of \textit{the answer-image} is shown in Fig.\ref{fig:5} (f) and displays a standard deviation of 0.0371. 
These standard deviation are much smaller than those between \textit{the answer-image} and \textit{the imitated-image}, 0.145.\\
　In most tested cases, the RL algorithm without regularization retrieved satisfactory images after an appropriate number of iterations. 
However, sometimes the RL algorithm did not converge to a suitable solution. 
Fig.\ref{fig:23} shows the standard deviation between \textit{the answer-image} and the deconvolved image as a function of iteration number; the solid curve shows the result of the original RL algorithm, and the dashed curve shows that of the RL algorithm with regularization using a $\lambda_{TM}$ of 0.0016. 
In the RL algorithm without regularization, the minimum standard deviation (the best output) occurred at the 29th iteration. 
For additional iterations, the standard deviation increases again, i.e. noise is enhanced. 
Since the optimal iteration number is a priori unknown, it is easy to continue iterating too long and hence enhance noise unduly. 
One example of this is seen in Fig.\ref{fig:5} (c) with 50 iterations. 
The RL algorithm with regularization is better behaved in this respect. Although, the standard deviation increased slightly with increasing number of iterations beyond the optimal iteration number (36th iteration in this case), this method still attained high accuracy without excessive noise enhancement long after the optimal iteration number. Whereas both methods, original RL algorithm and RL algorithm with regularization, attain almost the same accuracy at their optimal iteration number, the latter algorithm can provide a stable output regardless of the iteration number. \\
　As mentioned previously, although users need to assign a value to the regularization parameter $\lambda_{TM}$, it is difficult to find the appropriate one. 
Thus, we checked the accuracy of the various deconvolutions, adopting different values of $\lambda_{TM}$ to determine which parameter value provides the most reliable image. 
To visualize the correlations between \textit{the answer-image} and retrieved image, we used various log$_{10} \lambda_{TM}$ from $-2$ to $-5$ with a spacing of 0.1.  
In Fig.\ref{fig:22}, we plotted the corresponding standard deviations between \textit{the answer-image} and the image retrieved by the regularized RL algorithm and found that the minimum value (i.e., the best correlation between their images) is located around 0.0016 (equivalently $-2.8$ on the logarithmic scale). 
Therefore, this value of $\lambda_{TM}$ was adopted because it provided the best estimation of \textit{the answer-image}. 
Besides, Figs.\ref{fig:23} and \ref{fig:22} reveal that the best results obtained by the two versions of the RL algorithm are nearly of the same quality, although the algorithm in its regularized form leads to slightly lower standard deviations. The main shortcoming of the RL algorithm with regularization is illustrated by Fig.\ref{fig:22}. The appropriate value of $\lambda_{TM}$ to use for a particular type of data set is not {\it a priori} known and simulations cannot always be used to fix it, as we have done here.

\subsection{Observational results}
\subsubsection{Spatial distribution of granulation}
　Figure \ref{fig:7} shows the snapshot of the spatial distribution of the continuum intensity, as well as the convective velocity and the 5-min oscillations at a bisector level of 0.70. 
The top panels are derived from the observations before deconvolution, and the bottom panels correspond to the spatially deconvolved results. 
The deconvolution process enhances the continuum intensity contrast, from 7.2\% before applying the technique to 13.0\% after, which is close to the 14.4\% contrast found by \citet{Danilovic2008} from MURaM undegraded simulations. 
The remaining difference is partly due to the binning to the instrumental pixel scale, with the rest possibly due to a small defocus of the SP (see \citealt{Danilovic2008}). 
While we do not find notable differences in the shapes of the granules, we found the presence of small-scale spatial variations inside them that were very poorly visible before the deconvolution process. 
One example can be found around [1.0,1.0]~Mm: 
in the top panel (before deconvolution), the granules is nearly monotonously bright; in the bottom panel (after deconvolution), the darkening of the inner part of the granule can be seen clearly. 
A very weak indication of this darkening is already present in the original image, but is not nearly as clear as in the deconvolved one. \\
　The deconvolution process also improves the contrast of the convective velocity along with a much smaller enhancement of the 5-minutes oscillation.
The RMS of 0.61 km/s in the originally observed convective velocity field increased to 1.12 km/s after the deconvolution. 
Importantly, the downflows display a larger increase than the upflows. 
It can be seen immediately that the area occupied by downflows is larger in the deconvolved observation (bottom panel) than in the original observation (top panel). 
More quantitatively, we defined the area fraction of downflows as the ratio between the downflowing portion of the image area to the total area, and found that this value increased from 0.34 before the deconvolution to 0.48. 
To obtained these area fractions we discriminate between up- and downflows beyond a threshold of $\mp$0.18 km/s, respectively, and this value is the estimated error of the wavelength, using the spatially-time averaged profile of the Fe~{\sc i} 630.15~nm line to match the well-calibrated quiet-sun line profile of \citealt{Allende1998} (see more details in \citealt{Oba2017}).
This increase in the area fraction is because the weak upflows at the edge of each granule transformed into downflows after deconvolution, and the narrow intergranular channels sandwiched between granules appeared to increase in width. 
One example can be detected at around [0.3, 5.5]~Mm, where the deconvolved observation shows a clear downflow in a spatial location with zero LOS velocities in the original observation. \\
　Figure \ref{fig:14} displays scatter plots of the deconvolved versus original intensity (top panel) and LOS velocity (bottom panel). 
The slopes of the scatter plot for the continuum intensity and convective velocity field are 1.78 and 2.14, respectively, implying that the continuum intensity contrast and the convective velocity field are nearly/more than doubled by the deconvolution, respectively. 
In the case of the continuum intensity, the scatter is uniform for the whole range of intensities, whereas for the LOS velocities we found larger deviations for stronger downflows.

\subsubsection{Magnitude of convective velocity field}
The histogram of the convective velocity field in Fig.\ref{fig:24} shows quantitatively how deconvolution changes the upflows and downflows by different amounts.
These velocities were computed using bisector analysis (see Section 2.4 for more details) for the pixels in the center of the FOV along the scanning direction and around the central part of the slit, thus avoiding uncertainly deconvolved pixels close to the edges, as these need information outside the FOV. 
The magnitude of the convective velocity field at all the heights is enhanced in the deconvolved observation, especially for the downward motions. 
Before the deconvolution, at a bisector level of 0.70 (depicted by the black line in Fig.\ref{fig:24}), the distribution stretches from -2.0 (upward) to +1.5 (downward) km/s.   
Consequently, the blue-shifted signal is typically stronger than the red-shifted signal. 
After the deconvolution, downflows reach approximately the same magnitude as upflows with the amplitude ranging from -3.0 to +3.0 km/s. 
Similarly, the average speeds of upflows and downflows are found to be -0.64 km/s and +0.49 km/s before deconvolution, but -1.14 km/s and +1.20 km/s after that. \\
　The velocity histograms change their shape significantly as the height decreases. 
In the highest layer (blue line), the deconvolution process broadens the velocity distribution without significantly shifting the peak. 
In the lowest layer (black line), the deconvolution not only broadens the velocity distribution, but by changing the shape of the distribution, also shifts the peak. 
The first-moment of the convective velocity distribution (i.e., averaged over all pixels) is a rough indicator of how well up- and downflowing mass flux balance, although the density of gas also contributes and the difference in height of formation in granules and intergranular lanes affects the results as well, as shown in Fig.\ref{fig:25}.
Clearly, after the deconvolution, the average flow speeds lie closer to 0 km/s (the values fluctuate between +0.05 and -0.05 km/s), whereas prior to deconvolution, the average flow speed exceeded -0.2 km/s at the some bisector levels. 
At lower intensity levels (geometrically higher layers), this shift by the deconvolution process decreases (i.e., the average flow speeds before and after deconvolution is similar).

\subsubsection{Comparison with MHD simulations}
　For a direct comparison to the numerical simulations, we applied bisector analysis to the spectral profiles synthesized with the SPINOR code in the atmosphere computed by the MURaM code. 
We take this approach instead of comparing directly with the velocities in the simulated cubes, because the Doppler velocity fields obtained via bisector analysis are unavoidably smeared by the different atmospheric contributions to the line profile along the LOS and thus are smaller than the actual atmospheric velocity fields. 
We started with the original resolution simulations and computed the line profiles and bisectors. 
Then we spatially degraded them using the \textit{Hinode}/SP PSF. 
Histograms are displayed in Fig.\ref{fig:13}. 
It shows the same quantities as Fig.\ref{fig:24}, but now for velocities derived from the simulated histograms, both at original resolution and after convolution with the \textit{Hinode}/SP PSF. 
The RMS of the convective velocity field at a bisector level of 0.70, dropped from 1.10 km/s before the spatial convolution to 0.62 km/s after it, which is almost the same amplitude as the observation. 
The first moment of the velocity also shows the same trend as the observations. 
It became less negative i.e. tended more towards a red-shift through the spatial convolution. 
Thus its value goes from -0.11 km/s to -0.33 km/s. 
Note that these values are more blue-shifted than the observation. 
The area fraction of downflows decreases from 0.46 before the spatial convolution to 0.27 after that; averaged speeds of upflows and downflows are -1.06 km/s and +0.98 km/s before convolution, getting reduced to -0.75 km/s and +0.51 km/s after that. 
These up- and downflows are distinguished by the same threshold ($\mp$0.18 km/s) as the observational one (see Section. 3.2.1).}

\section{Discussions}
　The continuum contrast of \textit{Hinode}/SP is 7.2\% for the original observation and 13.0~\% after the deconvolution. 
We considered whether these values are realistic by comparing them with the results obtained using hydrodynamic simulations. The contrast of the synthesized image in the original pixel sampling ($\approx$10 km), the summed sampling ($\approx$100 km), and the summed sampling after the convolution with the instrument's PSF shows 16.8\%, 16.1\%, and 8.2\%, respectively. The third one is similar to the one observed by \textit{Hinode}/SP, i.e., 7.2\%, although the \textit{Hinode} value  is somewhat lower. 
According to \citet{Danilovic2008} this difference may be due to a small defocus in the \textit{Hinode}/SP, as the best focus is generally determined using the filtergraph.
The contrast value of 13.0\% in the deconvolved observation is lower than 16.1\% in the original simulation. 
This suggests that the image is not over-restored, so that the deconvolution provided a realistic image in terms of the contrast. 
Moreover, these values are consistent with the previous work of \citet{Danilovic2008}, who found RMS contrasts of 14.4\%, 7.5\%, and 7.0\% in the original pixel sampling ($\approx$10 km), the summed sampling  ($\approx$100 km) with spatial convolution, and the \textit{Hinode}/SP observation used in their work, respectively. 
Their contrast values are slightly lower than ours because their simulation included magnetic flux, which hinders the granulation and, hence, reduces the intensity contrast (\citealt{Biermann1941, Riethmuller2014}). 
The contrast of 13.0\% found in our work is also consistent with the value of 12.8\% estimated by the observational approach of \citet{Mathew2009}, who deconvolved \textit{Hinode} images using a PSF of the \textit{Hinode}/SOT derived using images during the Mercury transit. \\
　Before the deconvolution, the results displayed stronger convective upflows and weaker downflows. 
However, after the deconvolution, these amplitudes of up- and downflows became comparable, resulting from a preferential enhancement of the downflows. 
This enhancement can be explained by considering the difference in emergent intensity in granules and intergranular lanes. 
Convective blue- and red-shifted signals within a sub-arcsec spatial scale are unavoidably mixed because of the imperfect imaging performance of a telescope. 
Thus, they partially cancel each other, leading to spatially degraded signals. 
The higher intensities of granules compared to intergranular lanes, implied that more light from the brighter granules is scattered into the darker intergranular lanes than the other way around. 
Since radiation from granules is blue-shifted, it reduces the redshift of the radiation from intergranular lanes. 
This effect is larger than the decrease in the blueshift of radiation coming from granules, due to the smaller amount of straylight from intergranules, so that finally the downflows are decreased more strongly than the upflows. 
After deconvolution, the two velocities are better separated from each other again, so that downflows are more strongly enhanced than upflows.
The preferential enhancement in downflows by the deconvolution leads to some weak upflows (mainly at the boundaries of granules) turning into downflows after deconvolution. This either decreases the areas of particular granules somewhat, or brings to light, usually narrow downflow lanes that are not visible in the original data. These effects increase the area covered by downflows at the cost of that covered by upflows. \\
　Before the deconvolution, we found stronger upflows and weaker downflows, in agreement with previous works (\citealt{Yu2011b, Jin2009, Oba2017}).
After deconvolution, there was still an asymmetry between up- and downflows, but now with downflows being on average faster (although still covering a somewhat smaller area). 
The LOS flow averaged over the full FOV speed now lay within $\pm$0.05 km/s at all bisector levels, compared with an offset of up to -0.2 km/s in the original data. 
We found that the effect of the deconvolution on the data is similar in magnitude (but of opposite sign) as the changes introduced by convolving the data from the the MHD simulations.
In Fig.\ref{fig:24} for the observation and Fig.\ref{fig:13} for the numerical simulation, small differences, e.g., in the exact shape of the velocity histograms, are present between simulations and observations, but these are minor compared with the otherwise surprisingly good agreement. 
The consistency between the observations and the numerical simulation are a good indication that \textit{Hinode} resolves much of the granular structure. 
It also implies that the MURaM simulations provide a good description of solar granulation. \\
　We found that the magnitude of convective fields decreases with increasing height (see Fig.\ref{fig:24}), which is consistent with overshooting convection. 
The photosphere is convectively stable, so that buoyancy acts against the material's movement, and thus weakens the convective motions \citep{Stix2004}. 
We also detected that the effect of the deconvolution process on the convective motions is height-dependent; 
at the higher layers, the changes through the deconvolution process are less prominent. 
We believe that the reason behind this behavior is the photospheric temperature stratification, in which, towards higher layers, the intensity contrast in the granulation becomes smaller and eventually reverses, i.e.,  \textit{reversed granulation} is seen. 
In this layer, dark blue-shifted granules are surrounded by bright red-shifted intergranular lanes \citep{Kostik2009}. 
The transition from the normal granulation to the reversed one occurs at roughly 130 to 140~km according to the numerical simulations of \citet{Wedemeyer2004} and \citet{Cheung2007} and at roughly 140~km in the observational work of \citet{RuizCobo1996}. 
The lowest bisector level investigated the present work samples an average height (135 km) roughly corresponding to the height of this transition. 
At around this geometrical height, the first-moments before and after the deconvolution process show no significant difference, which is consistent with this temperature distribution because the intensity from granules and intergranular lanes should be almost identical.

\section{Summary}
　To correct the instrumental imaging performance, we applied a deconvolution technique based on the Richardson-Lucy algorithm to \textit{Hinode}/SP data. 
We incorporated a regularization term into the algorithm, which minimizes the risk of enhancing the noise component. 
Using a synthesized image, we confirmed that the RL algorithm in a regularized form restricted the noise enhancement. 
The deconvolution of the data led to a major change in the magnitude of the LOS velocity obtained via bisector analysis, in particular at higher intensities in the spectral line. 
The resulting convective velocity amplitudes are almost the same for up- and downflows (ranging from approximately -3.0 km/s to 3.0 km/s at bisector level 0.70). 
Whereas the original data showed a global upflow (i.e., net LOS velocity averaged over all pixels) of up to -0.2 km/s, after the deconvolution this value dropped to below $\pm$0.05 km/s. 
This indicates that the downflows were preferentially increased by the deconvolution, while the upflows were only moderately enhanced. 
The upflow obtained from the original data is consistent with the earlier investigations based on \textit{Hinode} data (\citealt{Yu2011b}, \citealt{Jin2009}, and \citealt{Oba2017}). 
We found the observed LOS velocities before and after deconvolution matched well with those derived from the numerical simulations, obtained by applying bisector analysis to the synthesized spectral profiles after and before convolution with the \textit{Hinode} PSF, respectively. 
This agreement with MHD simulations is very heartening and suggests that the best current MHD simulations provide a relatively good description of solar granulation. 
Some remaining disagreement in the RMS contrast and velocity distributions may be due to a residual defocus in the \textit{Hinode} data. 
Observations at higher spatial resolution (e.g. with the Sunrise balloon-borne solar observatory; \citealt{Solanki2010, Solanki2017, Barthol2011}) and covering more spectral lines will provide a more sensitive test. \\
　The deconvolution also changed the spatial distribution of the velocity field, such that the area occupied by upflowing material at the granular edges was trimmed off by downflows pertaining to the intergranular lanes. 
By revealing more downflow lanes, a deconvolution technique such as ours will provide new insights into magneto-convection by allowing to see the immediate surroundings of small flux tubes, preferentially located in intergranular lanes (\citealt{Title1987, Grossmann-Doerth1988, Solanki1989}). 
This should enable the magnetic features to be better probed, e.g., as done by \citet{Dominguez2003} and \citet{Buehler2015} using complementary approaches. 
Possible future applications of deconvolution include catching the detailed process of convective collapse coinciding with a downflow within the magnetic features \citep{Parker1978}, as observed by \citet{Nagata2008} and \citet{Requerey2014}, and described using MHD simulations by \citet{Danilovic2010}, or the magneto-acoustic waves generated intermittently in flux tubes (e.g., \citealt{Stangalini2014}, \citealt{Jess2016}), which are excited by the surrounding convective motions \citep{Kato2016}.

\acknowledgments
\textit{Hinode} is a Japanese mission developed and launched by ISAS/JAXA, collaborating with NAOJ as a domestic partner, NASA and STFC (UK) as international partners. Scientific operation of the \textit{Hinode} mission is conducted by the \textit{Hinode} science team organized at ISAS/JAXA. This team mainly consists of scientists from institutes in the partner countries. Support for the post-launch operation is provided by JAXA and NAOJ (Japan), STFC (U.K.), NASA, ESA, and NSC (Norway). We are grateful to the \textit{Hinode} team for performing observations on August 25, 2009, which were well suited to this analysis. 
The present study was supported by the Advanced Research Course Program of SOKENDAI and by JSPS KAKENHI Grant Number JP16J07106. 
This project has received funding from the European Research Council (ERC) under the European Union's Horizon 2020 research and innovation programme (grant agreement No 695075) and has been supported by the BK21 plus program through the National Research Foundation (NRF) funded by the Ministry of Education of Korea.

\bibliographystyle{apj}
\bibliography{myrefs}

\clearpage

\begin{figure}
\includegraphics[width=12cm]{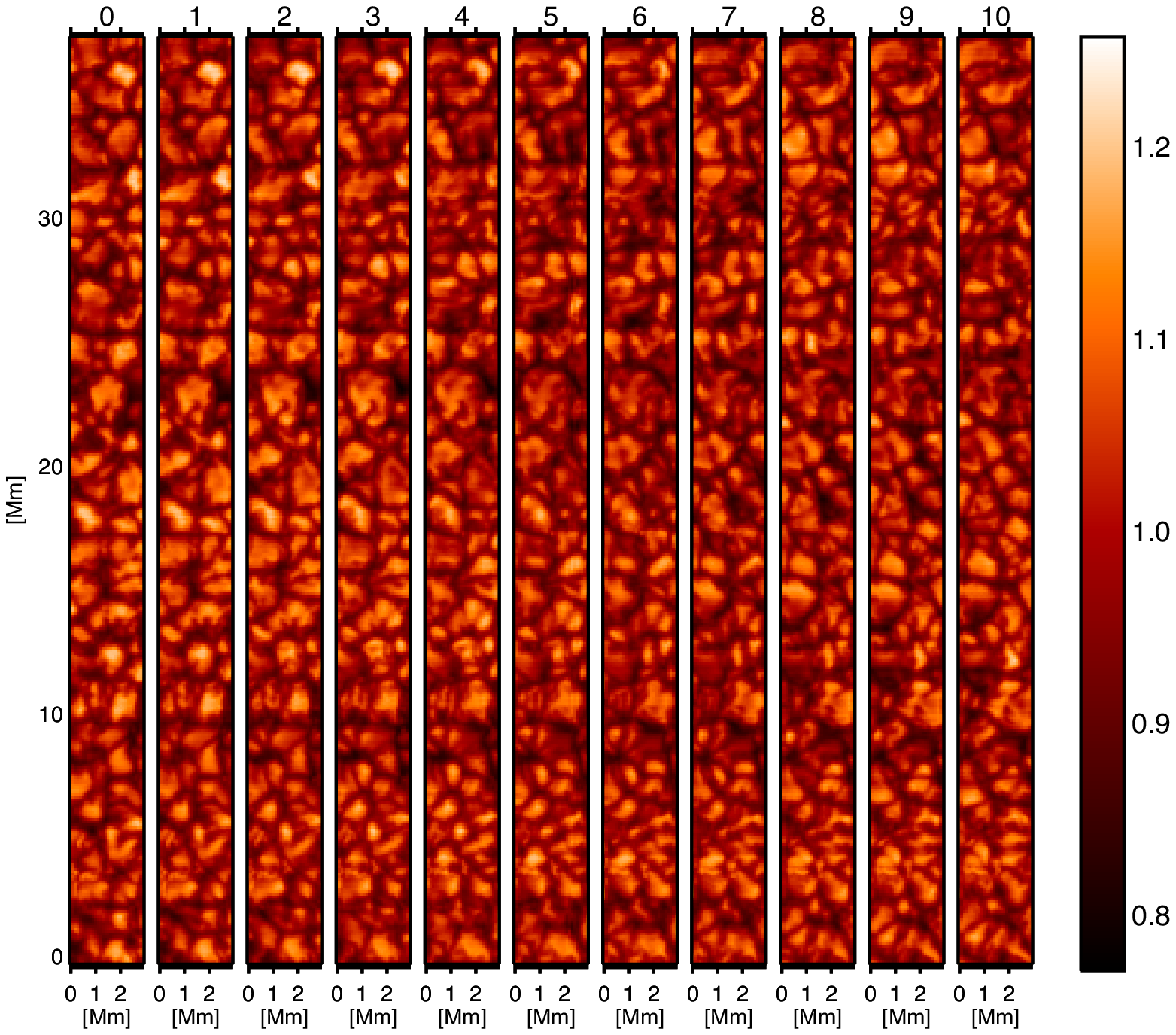}
\caption{Evolution of the observed continuum intensity. The vertical axis corresponds to the slit-direction, and the horizontal axis represents the scanning-direction. The spatial sampling is roughly 100 km and the numbers above each map designate the time sequence with a cadence of 62~s.}
\label{fig:1}
\end{figure}

\begin{figure}
\includegraphics[width=12cm]{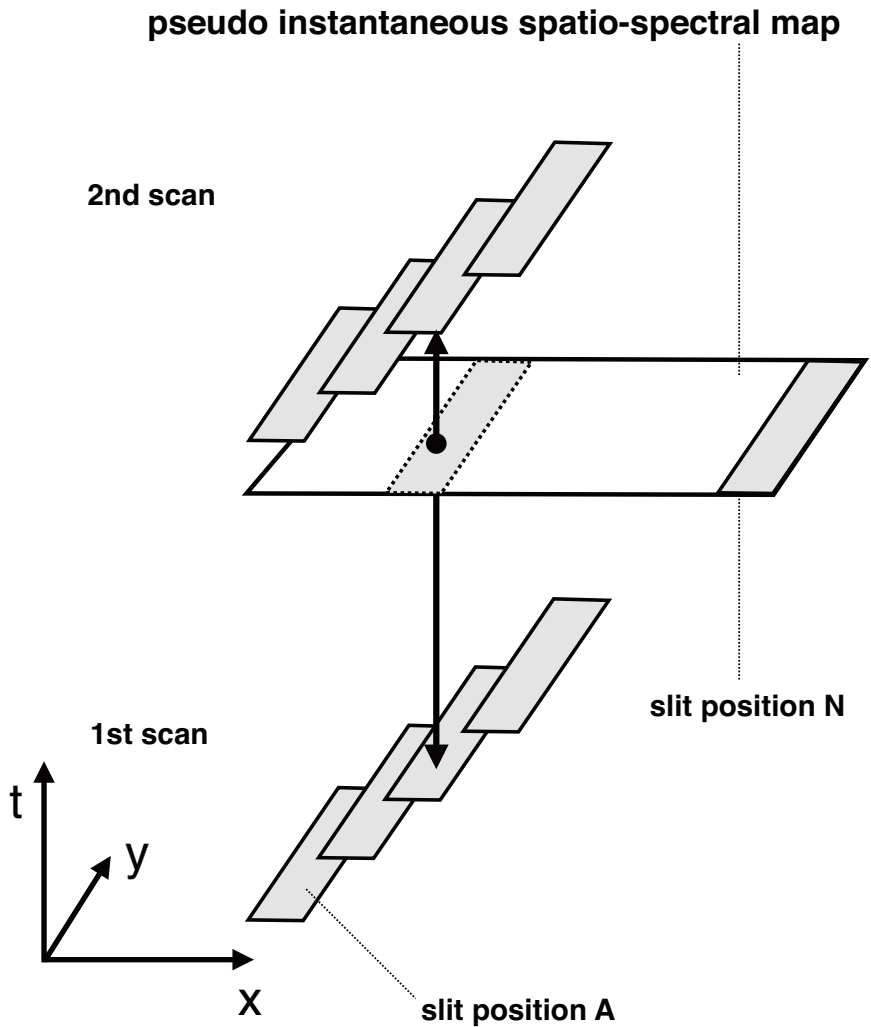}
\caption{Illustration of the making of a pseudo-instantneous two-dimensional spectral map.
Spectra are recorded along the slit at each x-position, starting from the left (at position A) and moving to the right with time, i.e., scanning along the x-direction.
After the last slit position (position N) of a scan, the scanning position goes back to the first position (position A) and the process is repeated. 
To make a pseudo-instantaneous spatio-spectral map at a given time at the end of the first scan (indicated by the larger rectangle), the slit data at each spatial x-position are computed by linearly interpolating between the 1st and 2nd scans, as indicated by the vertical double-headed arrow and the dotted rectangle lying in the larger rectangle. }
\label{fig:20}
\end{figure}


\begin{figure}
\includegraphics[width=13cm]{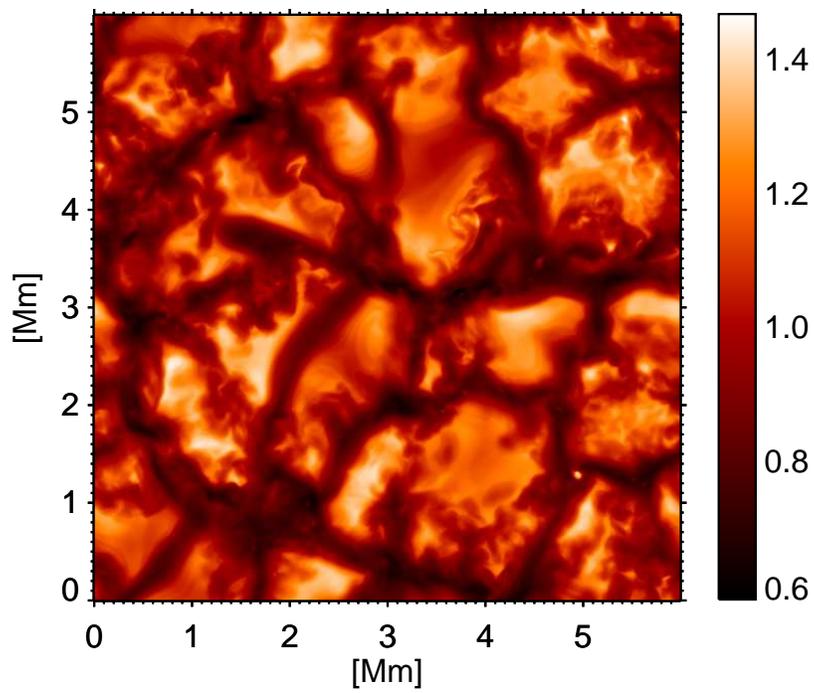}
\caption{Emergent continuum intensity map for the simulation box at one snapshot considered in this paper.}
\label{fig:21}
\end{figure}

\begin{figure}
\includegraphics[width=13cm]{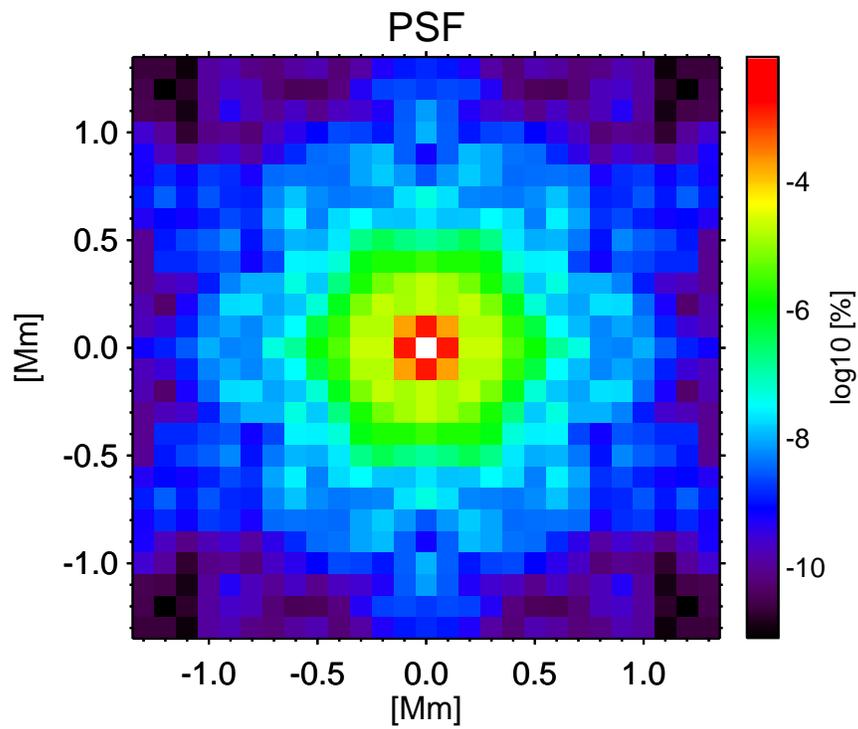}
\caption{PSF of \textit{Hinode}/SP provided by S. Danilovic and obtained from the \textit{Hinode}/SOT pupil noted in \citet{Suematsu2008}. One pixel corresponds to a spatial sampling of 100 km, the same as in the original \textit{Hinode} data used here.}
\label{fig:2}
\end{figure}

\begin{figure}
\includegraphics[width=13cm]{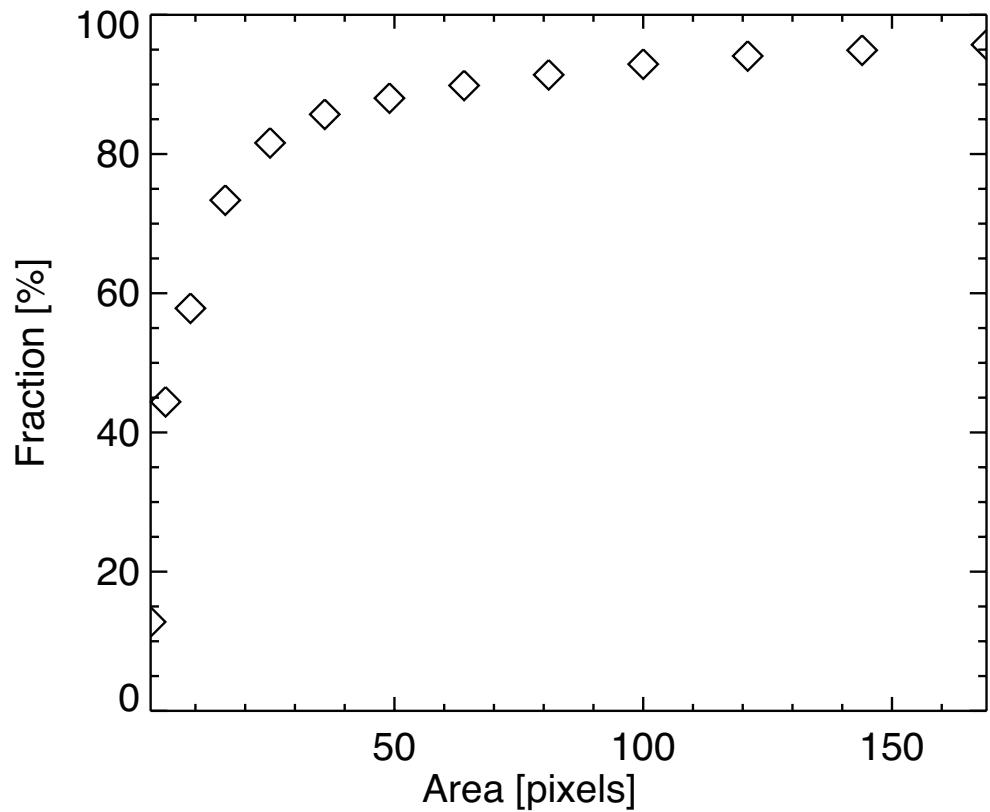}
\caption{Total fraction of the PSF contained within a given square area, expressed in pixels, i.e. fraction of the total energy within a given square area on the detector. 
Diamond symbols correspond to the fraction of the PSF contained within squares with an area of 1, 4, 9, 16, ..., 169 pixels (equivalent to a length of one side of the square of 1, 2, 3, 4, ..., 13 pixels). }
\label{fig:3}
\end{figure}

\begin{figure}
\includegraphics[width=12cm]{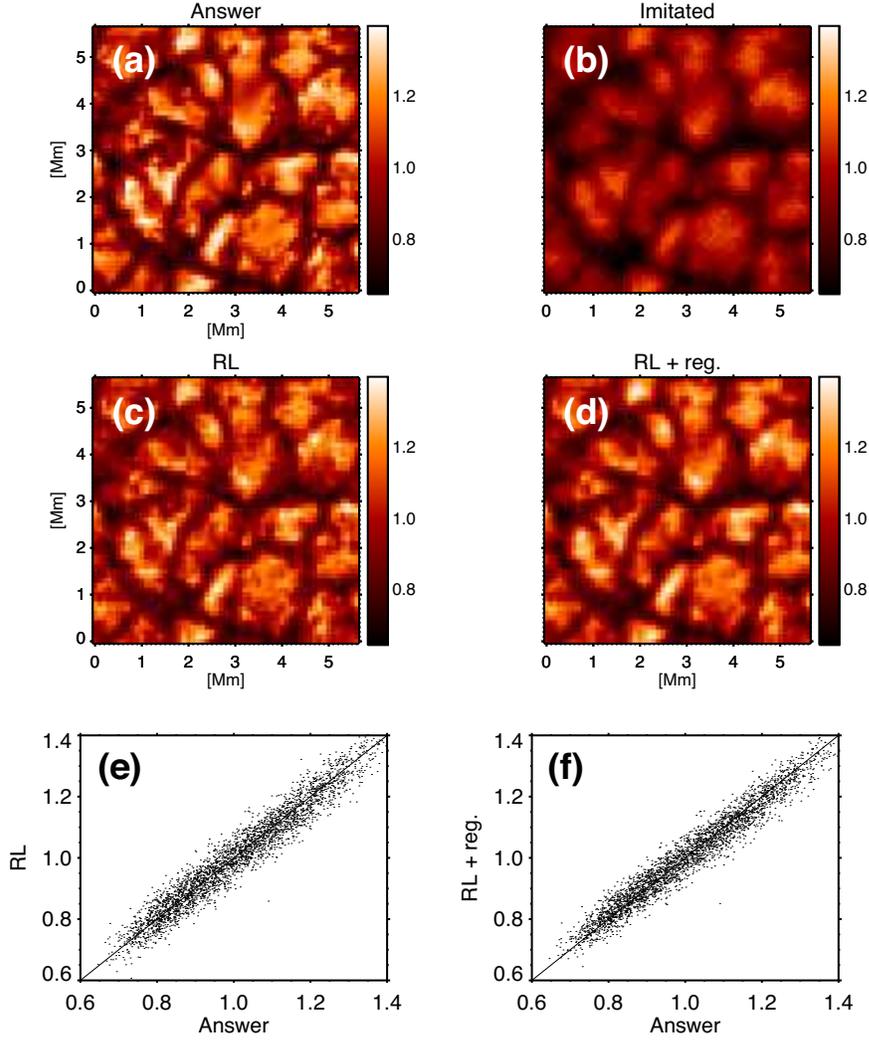}
\caption{Comparison of the original image to the one retrieved through deconvolution. 
\textit{Panel (a)}: \textit{the answer} image, which is the simulated original continuum intensity with a spatial pixelation of $\approx$ 100 km. 
\textit{Panel (b)}: \textit{the imitated} image, which is produced by convolving the answer-image by the PSF and additionally introducing Gaussian random noise. 
\textit{Panel (c)}: the intensity map retrieved by the RL algorithm without any regularization at an iteration number of 50. 
\textit{Panel (d)}: the intensity map retrieved by the extended RL algorithm including the regularization term. 
\textit{Panel (e)} is a scatter plot between \textit{the answer-image} (\textit{Panels (a)}) and the intensity retrieved through the RL algorithm without regularization (\textit{Panel (c)}). 
\textit{Panel (f)} depicts a scatter plot between \textit{the answer-image} (\textit{Panel (a)}) and the intensity retrieved through the RL algorithm with regularization (\textit{Panel (d)}).}
\label{fig:5}
\end{figure}

\begin{figure}
\includegraphics[width=13cm]{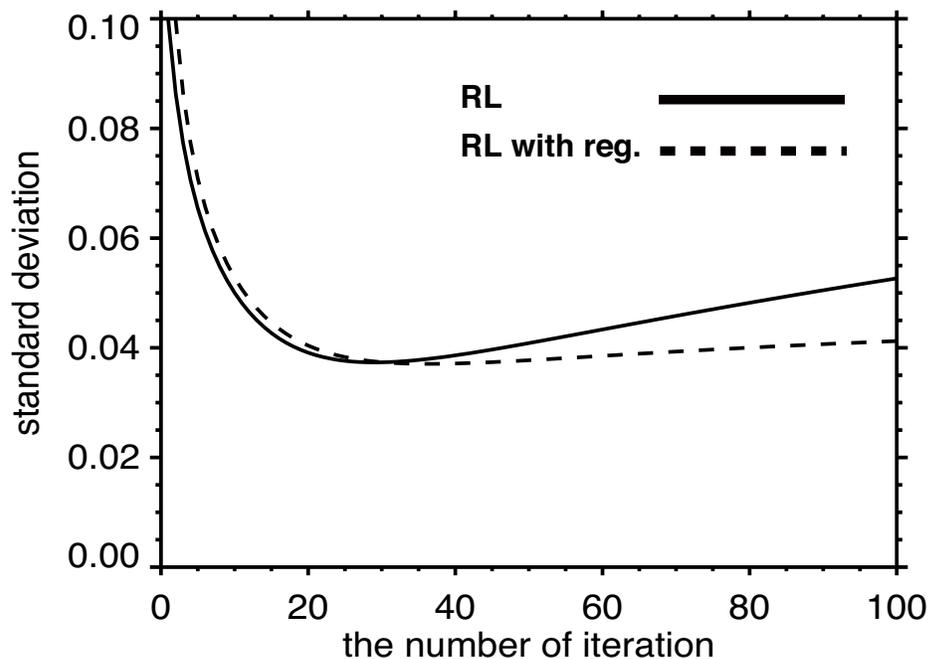}
\caption{Standard deviation between \textit{the answer-image} and the deconvolved image as a function of the iteration number.  
The solid line represents the standard deviation achieved by the original RL algorithm, and the dashed one that by the RL algorithm with regularization using $\lambda_{TM}$ = 0.016 (see main text and Fig.\ref{fig:22} for the reasons behind this choice). }
\label{fig:23}
\end{figure}

\begin{figure}
\includegraphics[width=13cm]{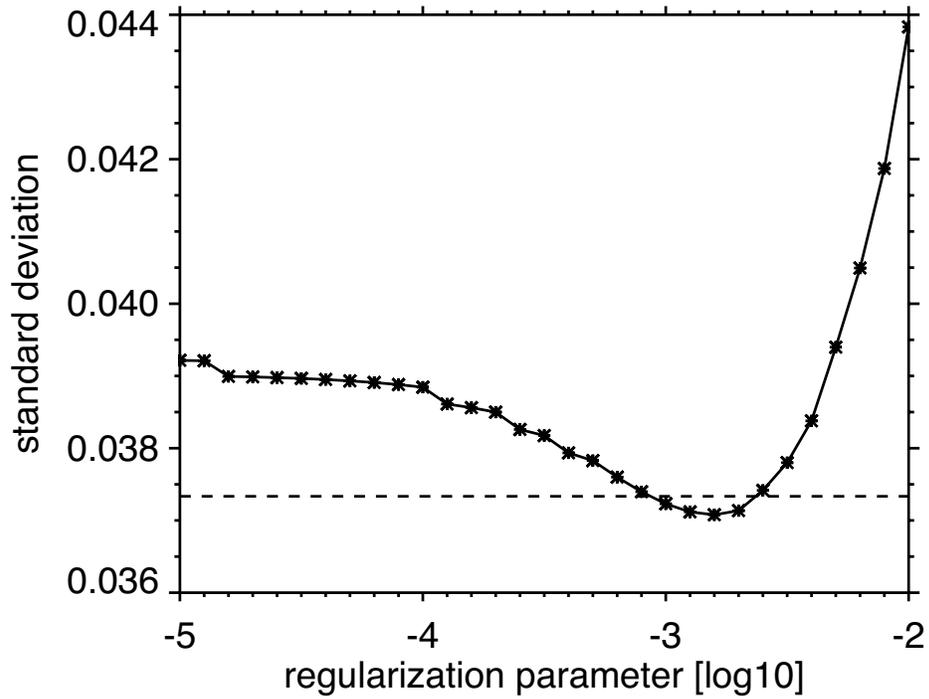}
\caption{Standard deviation between \textit{the answer-image} and the image retrieved by the RL algorithm with regularization as a function of the regularization parameter, $\lambda_{TM}$. 
The horizontal dashed line indicates the minimum standard deviation reached by the RL algorithm without regularization.}
\label{fig:22}
\end{figure}

\begin{figure}
\includegraphics[width=13cm]{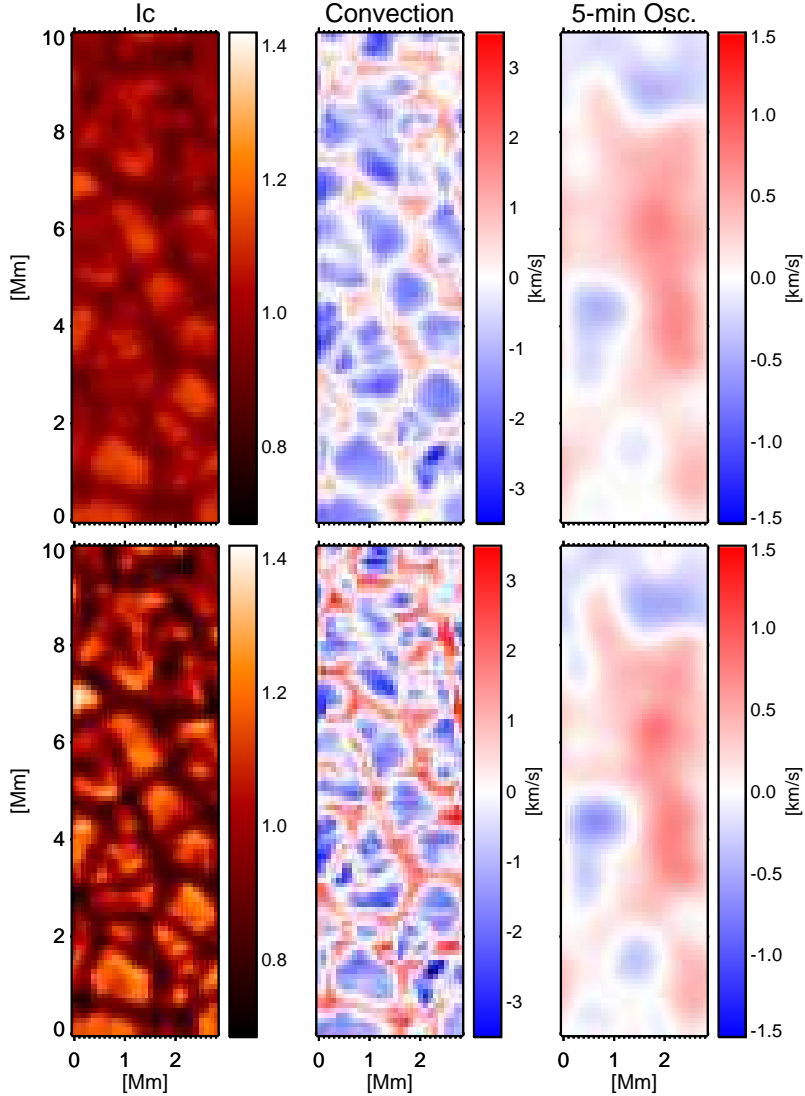}
\caption{
Granulation and p-mode oscillation images obtained from original and deconvolved \textit{Hinode}/SP data. \textit{The left column} shows the continuum intensity, \textit{the center column} displays the convective velocity field at a bisector level of 0.70, and \textit{the right column} depicts the 5-minute oscillation at the same level. \textit{The top row} corresponds to the observations before the deconvolution, while \textit{the bottom row} displays the deconvolved ones. Positive velocity signifies a red-shift (downward), a negative one a blue-shift (upward). }
\label{fig:7}
\end{figure}

\begin{figure}
\includegraphics[width=13cm]{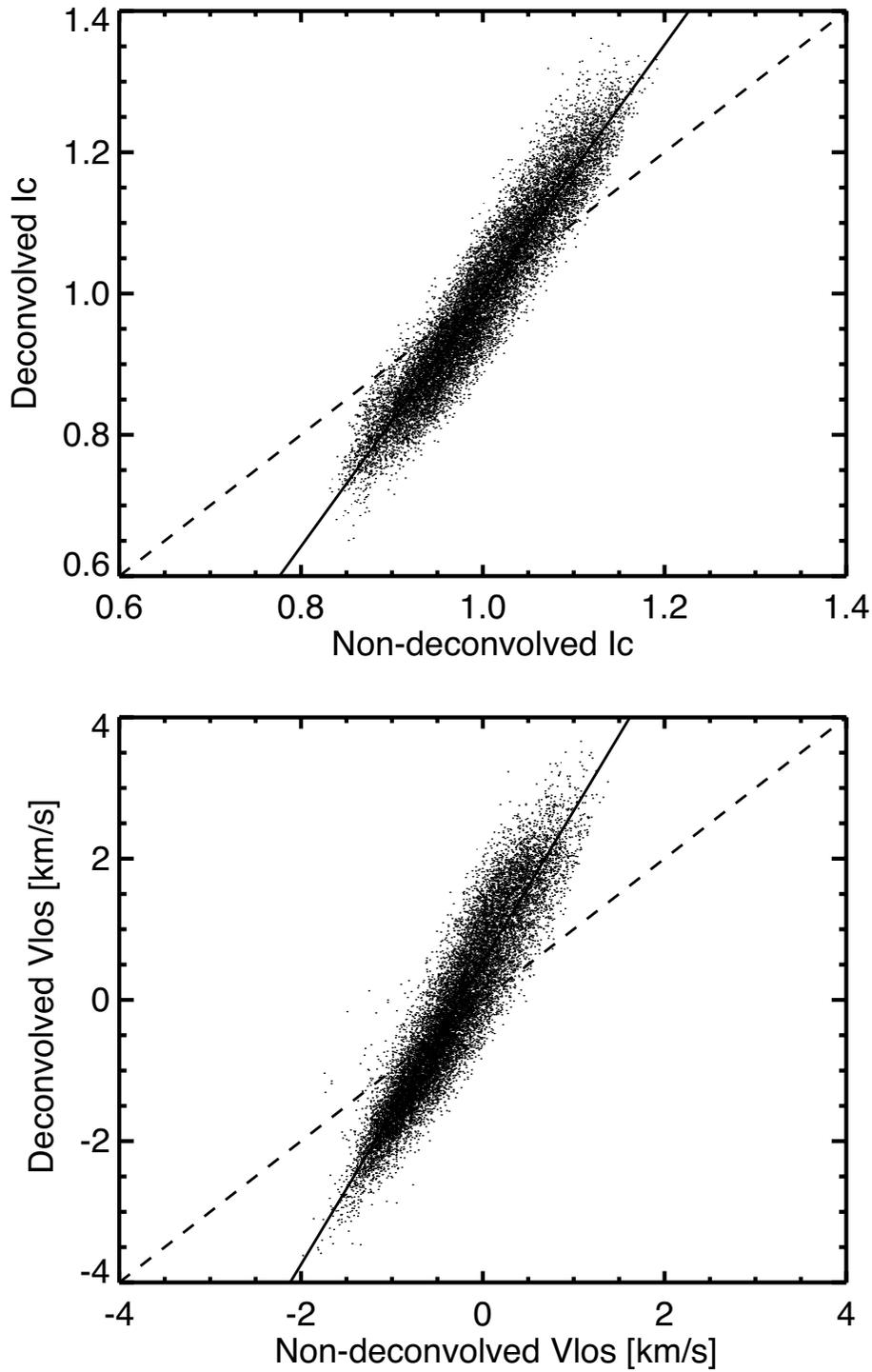}
\caption{Upper panel: Scatter plot of the continuum intensity after deconvolution vs. that before deconvolution. Lower panel: the same for the LOS component of the convective velocity field at a bisector level of 0.70. Solid lines corresponds to the best fitted line, and dashed lines indicate a slope of unity. }
\label{fig:14}
\end{figure}


\begin{figure}
\includegraphics[width=13cm]{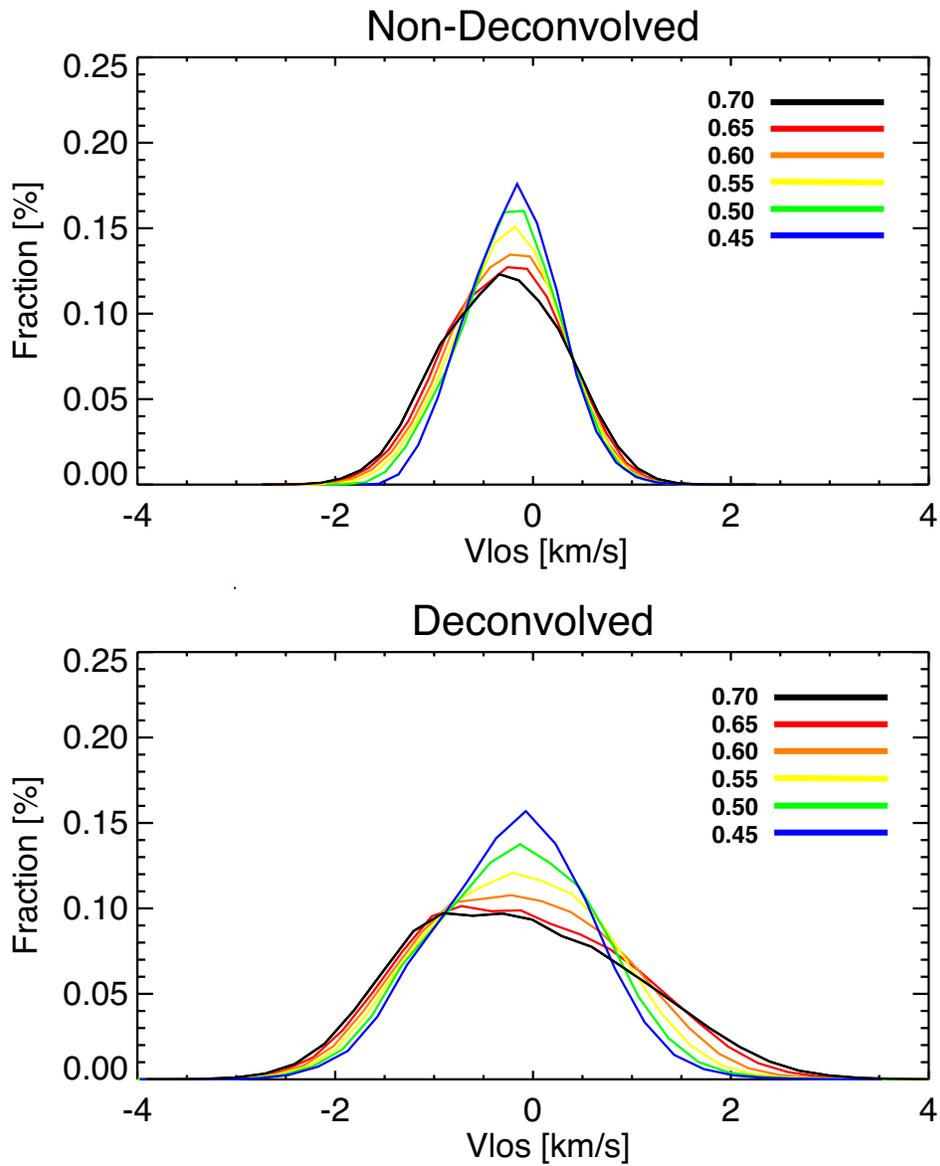}
\caption{Histogram of the convective velocity field before the deconvolution (top) and after the deconvolution (bottom). The sign of the velocity has the same meaning as in Fig.\ref{fig:14}. The colors represent different bisector levels, listed in the upper right corners of the frames. }
\label{fig:24}
\end{figure}

\begin{figure}
\includegraphics[width=13cm]{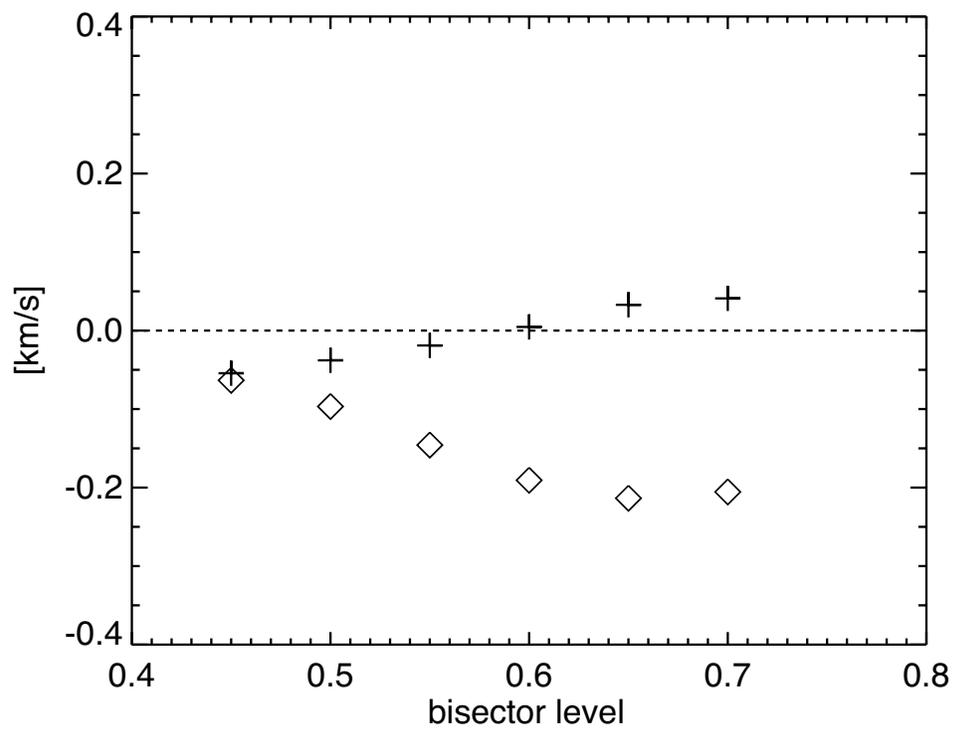}
\caption{First moment of the convective velocity fields at the 6 bisector levels. Diamonds indicate the first-momentum of the velocity distribution before the deconvolution, whereas crosses represent it after the deconvolution.}
\label{fig:25}
\end{figure}

\begin{figure}
\includegraphics[width=13cm]{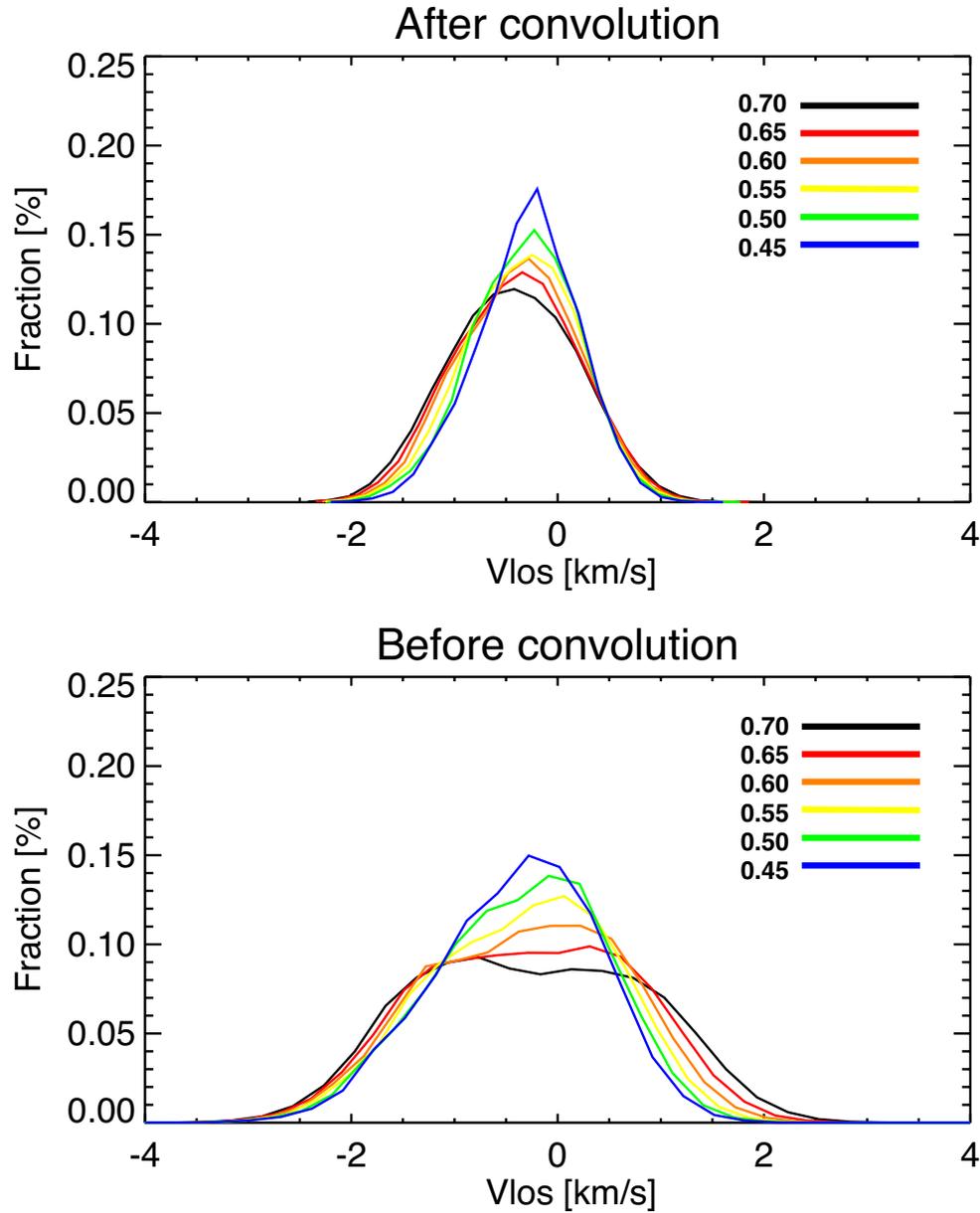}
\caption{Histogram of the LOS velocity, derived via bisector analysis from spectral lines synthesized in the MHD simulation. Upper panel: after the convolution with the \textit{Hinode} PSF. Bottom panel:  before the convolution. Positive values correspond to downflows and negative values mean upflows. 
The colors depict the different bisector levels identified in the upper right corners of the frames. These levels are the same as in Fig.\ref{fig:24}. } 
\label{fig:13}
\end{figure}

\end{document}